\documentclass[aps,prb,reprint,amsmath,amssymb,floatfix,superscriptaddress]{revtex4-2}

\usepackage{graphicx,xcolor,multirow,bm,hyperref}

\usepackage[ignoreunlbld,norefs,nocites]{refcheck}

\begin{document}

\title{Correlation energy of the spin-polarized electron liquid by quantum Monte Carlo}

\author{Sam Azadi}

\email{sam.azadi@physics.ox.ac.uk}

\affiliation{Department of Physics, Clarendon Laboratory, University of Oxford, Parks Road, Oxford OX1 3PU, United Kingdom}
\affiliation{Department of Chemistry, Paderborn Center for Parallel Computing, Paderborn University, 33098 Paderborn, Germany}

\author{N.\ D.\ Drummond}

\affiliation{Department of Physics, Lancaster University, Lancaster
  LA1 4YB, United Kingdom}

\author{Sam\ M.\ Vinko}

\affiliation{Department of Physics, Clarendon Laboratory, University of Oxford, Parks Road, Oxford OX1 3PU, United Kingdom}
\affiliation{Central Laser Facility, STFC Rutherford Appleton Laboratory, Didcot OX11 0QX, United Kingdom}

\date{\today}

\begin{abstract}
Variational and diffusion quantum Monte Carlo (VMC and DMC) methods with Slater-Jastrow-backflow trial wave functions are used to study the spin-polarized three-dimensional uniform electron fluid.  We report ground state VMC and DMC energies in the density range $0.5 \leq r_\text{s} \leq 20$. Finite-size errors are corrected using canonical-ensemble twist-averaged boundary conditions and extrapolation of the twist-averaged energy per particle calculated at three system sizes ($N=113$, 259, and 387) to the thermodynamic limit of infinite system size. The DMC energies in the thermodynamic limit are used to parameterize a local spin density approximation correlation function for inhomogeneous electron systems.
\end{abstract}

\maketitle

\textit{Introduction.} The three-dimensional uniform electron liquid (UEL) is an important model for studying many-body interactions in fermionic systems and for describing the electronic structures of real materials \cite{Pines66,Ceperley80,Giuliani05, Loos16,Dornheim16,Groth17,Dornheim22,Azadi22,Gino,Azadi23}. The UEL model at high density can be used to understand the behavior of real systems under extreme conditions, for example in studies of warm dense matter, which is an exotic, highly compressed state of matter that exists between solid and plasma phases at high temperatures and pressures \cite{Dornheim16,Dornheim18}. Furthermore, accurate calculations of the correlation energies of spin-polarized electron liquids are essential for density functional theory (DFT) as they enable the parameterization of spin exchange-correlation functionals that allow DFT to be used to study the magnetic properties of materials. 

The pairwise Coulomb interaction introduces many-body correlations in fermionic systems such as the electron liquid. The correlation energy of the electron liquid is defined as the difference between the exact ground-state energy per particle and the Hartree-Fock (HF) ground-state energy per particle. The correlation energy is only a small percentage of the total energy of an electronic system, but it significantly affects the chemical bonding, electronic structure, and magnetic properties of materials \cite{Lowdin55,Jones15,Dreizler}. The Kohn-Sham formalism of DFT \cite{Hohenberg64,Kohn65}, which is widely used to describe the chemical and physical properties of real systems, depends on the correlation energy of the electron liquid \cite{Perdew81, Perdew92, Vosko80} as calculated by many-body wave function-based methods \cite{Sun10,Loos11,Bhattarai18, Gell57}, including quantum Monte Carlo (QMC) simulation \cite{Ceperley80,Holzmann20, Ceperley77,Ceperley78,Azadi21,Spink13,Shepherd12b,Ruggeri18}. The variational (VMC) and diffusion quantum Monte Carlo (DMC) methods \cite{Ceperley80,Foulkes01} used in this work are stochastic methods for determining ground state expectation values of quantum operators. 

This work presents new VMC and DMC results for the correlation energy of the ferromagnetic (i.e., spin-polarized) three-dimensional UEL (3D-UEL), which are lower than previously reported results.  QMC energies obtained in finite simulation cells obey the variational principle, and it is reasonable to assume that the QMC energy per particle extrapolated to the infinite system-size limit is also in practice an upper bound on the true energy per particle.  Hence the fact that our energies are lower than previous works suggests that our results are more accurate.

We used the VMC and DMC techniques to obtain 3D-UEL ground state energies at high and intermediate densities ($0.5 \leq r_\text{s} \leq 20$). In the VMC method, parameters in trial wave functions are optimized according to the variational principle, with energy expectation values calculated by Monte Carlo integration in the $3N$-dimensional space of electron position vectors.  In the DMC method, the imaginary-time Schr\"{o}dinger equation is used to evolve a statistical ensemble of electronic configurations towards the ground state. Fermionic antisymmetry is maintained by the fixed-phase approximation, in which the complex argument of the wave function is constrained to equal that of an approximate trial wave function optimized within VMC\@. The \textsc{casino} package was used for all our QMC calculations \cite{casino}.
 
\textit{Trial wave function.} The simplest fermionic wave function is a Slater determinant of one-electron orbitals, which describes exchange effects but not correlation. Multideterminant wave functions \cite{Shepherd12} and pairing (geminal) wave functions \cite{Marchi09} can be used to describe correlation effects in electronic systems. However, the most efficient method of going beyond the Slater wave function in QMC calculations is to multiply it by a Jastrow factor $\exp(J)$, resulting in a Slater-Jastrow wave function \cite{Ceperley77,Ceperley78, Ceperley80}.  The Jastrow factor usually depends explicitly on the distances between particles, thereby allowing a very compact description of correlation effects with a relatively small number of variational parameters. The Jastrow factor is positive everywhere and symmetric with respect to the exchange of indistinguishable particles, so it does not change the nodal surface defined by the rest of the wave function.  By evaluating the orbitals in the Slater determinant at quasiparticle coordinates ${\bf X}({\bf R})$, which are functions of all the electron positions ${\bf R}$, we introduce a backflow transformation \cite{Kwon98,Pablo06}, and the resulting wave function is referred to as a Slater-Jastrow-backflow (SJB) wave function. We used a SJB trial spatial wave function $\Psi({\bf R})=e^{J({\bf R})}S({\bf X}({\bf R}))$ for all the systems studied, where ${\bf R}=({\bf r}_1,\ldots,{\bf r}_N)$ is the $3N$-dimensional vector of electron coordinates. The antisymmetric Slater part $S$ is a product of determinants of single-particle orbitals which are of the free-electron form $\psi_\mathbf{k}(\mathbf{r})=\exp(i\mathbf{k} \cdot \mathbf{r})$, where wavevector $\mathbf{k}$ is a reciprocal lattice vector of the simulation cell offset by twist vector $\mathbf{k}_\text{s}$, where $\mathbf{k}_\text{s}$ lies in the supercell Brillouin zone. The details of our Jastrow factor and backflow functions are provided in our previous works \cite{Azadi23,Azadi22,Drummond04}.

\textit{Finite-size effects.} Monte Carlo-sampled canonical-ensemble twist-averaged (TA) boundary conditions were used to reduce quasirandom single-particle finite-size (FS)
errors in total energies due to momentum quantization effects \cite{Lin01,Azadi15,Drummond08,Holzmann16,Azadi19}. The HF kinetic and exchange energies were used as control variates to improve the precision of the TA energy.  Systematic FS errors due to the use of the Ewald interaction rather than $1/r$ to evaluate the interaction between each electron and its exchange-correlation hole and the incomplete description of long-range two-body correlations were removed by fitting
\begin{equation} \bar{E}_\text{DMC}(N)=E(\infty)+b/N \label{eq:fs_extrap} \end{equation}
to the TA DMC energy per particle $\bar{E}_\text{DMC}(N)$ at different system sizes $N$, where $b$ and $E(\infty)$ are fitting parameters. Unlike the previous work of Spink \textit{et al.}\ \cite{Spink13}, we do not rely on analytic FS correction formulas \cite{Chiesa06,Drummond08}, but instead use the analytic results to provide the exponents used in FS extrapolation formulas.  All our calculations were performed using face-centered cubic (fcc) simulation cells, maximizing the distance between each particle and its closest periodic image.

\begin{figure}[!htbp]
    \centering
    \includegraphics[scale=0.4]{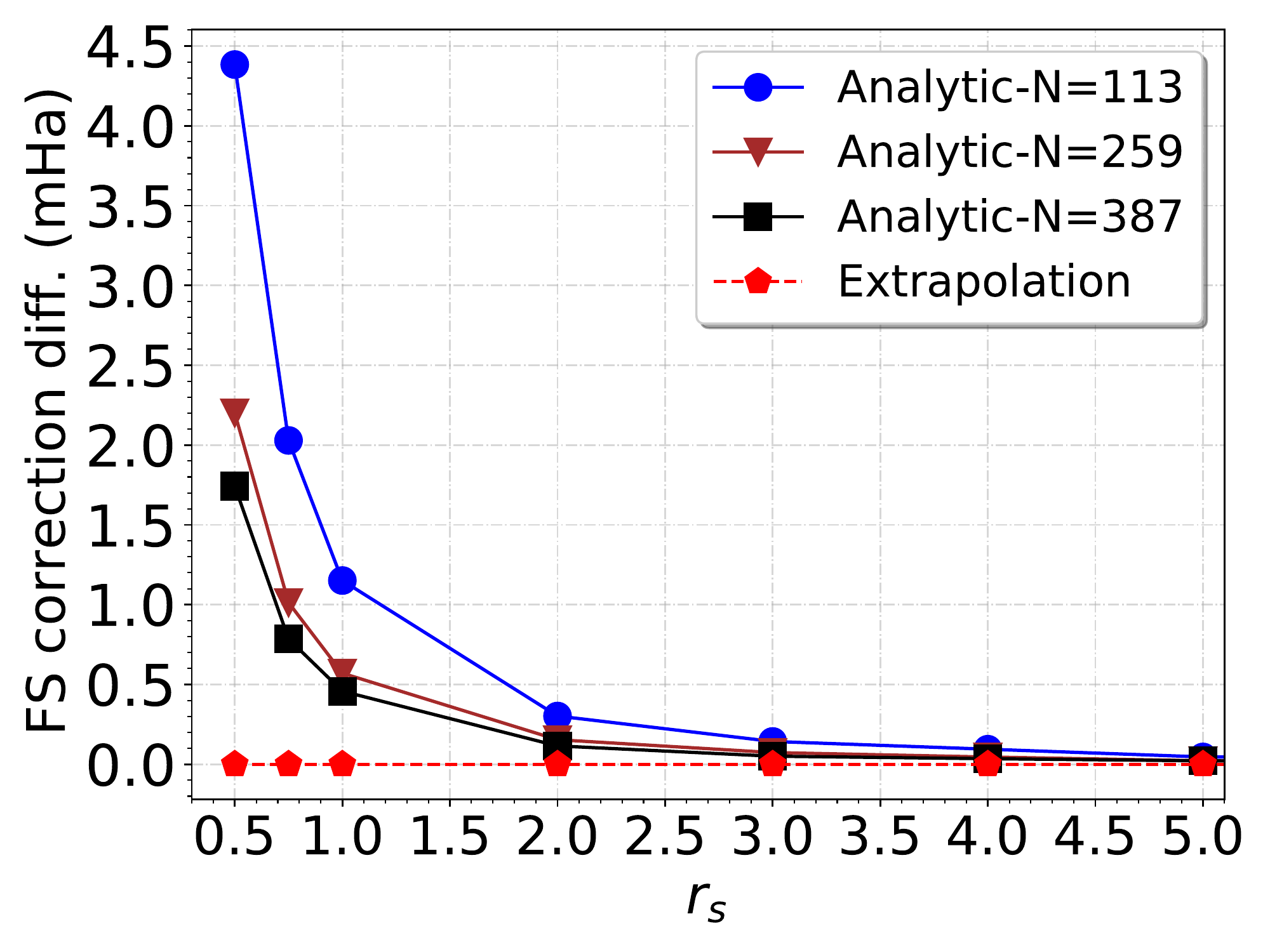}
    \caption{Difference between FS corrections obtained analytically and by extrapolation to infinite system size using Eq.\ (\ref{eq:fs_extrap}), plotted against density parameter $r_\text{s}$.} 
    \label{fig:FS}
\end{figure}

The leading-order analytical FS correction to the energy per electron of the 3D-UEL is $\Delta E_1=\frac{\sqrt{3/r_\text{s}^3}}{2N}$ \cite{Chiesa06}, where $N$ is the number of electrons. The next-to-leading-order correction is
\begin{equation}
    \Delta E_2 = \frac{C}{\pi r_\text{s}^2 (2N)^{4/3}\left[{(1+\zeta)}^{2/3}+{(1-\zeta)}^{2/3}\right]},
\end{equation}
where $\zeta$ is the spin polarization and $C=5.083$ for an fcc simulation cell \cite{Drummond08}. We calculated the analytical FS correction $\Delta E = \Delta E_1 + \Delta E_2 + \Delta E_\text{SP}$, where $\Delta E_\text{SP} = T_\text{HF}(\infty) - \bar{T}_\text{HF}(N)$ is the correction for residual single-particle errors in the TA HF kinetic energy $\bar{T}_\text{HF}$. Figure \ref{fig:FS} illustrates the difference between analytical FS corrections $\Delta E$ to the TA energies per particle and FS corrections obtained by extrapolation [i.e., $E_\text{DMC}(\infty) - \bar{E}_\text{DMC}(N)$, where $E_\text{DMC}(\infty)$ is the DMC energy at the infinite system size limit obtained by extrapolation using Eq.\ (\ref{eq:fs_extrap}) and $\bar{E}_\text{DMC}(N)$ is the TA DMC energy for a finite simulation cell containing $N$ electrons].
The analytical FS corrections do not include an $O(N^{-1})$ contribution arising from FS effects due to backflow \cite{Holzmann16}; however, these FS effects and any further $O(N^{-1})$ FS effects are removed by extrapolation. 

\textit{Quantum Monte Carlo energies.} We performed QMC calculations for the ferromagnetic 3D-UEL in simulation cells with $N=113$, 259, and 387 electrons for each density. Our DMC energies for $r_\text{s}=0.5$, 0.75, 1, 2, 3, 4, 5, 10, and 20 were obtained using time steps $\tau=0.005$, 0.01, 0.02, 0.05, 0.1, 0.16, 0.25, 1.0, and 2.0, respectively, with a target walker population of 2560. The time step bias for the selected $\tau$ is negligible \cite{Azadi22, Azadi23}. The TA QMC energies for different system sizes are listed in Table \ref{table:EvsN}.
\begingroup
\squeezetable
\begin{table*}[!htbp]
\centering
\caption{\label{table:EvsN} TA HF, VMC, and DMC total energies in Ha per electron of spin-polarized 3D-UELs. $N$ is the number of electrons in the fcc simulation cell. TA VMC energy variances in Ha$^2$ are also reported. }
 \begin{tabular}{l c c c c c c c c c}
 \multicolumn{10}{c}{$N=113$} \\ 
 \hline \hline
 $r_\text{s}$ & 0.5 & 0.75 & 1.0 & 2.0 & 3.0 & 4.0 & 5.0 & 10.0 & 20.0 \\ 
 \hline
  $\bar{E}_{\rm HF}$  &$5.834558(6)$&$2.330131(3)$&$1.162757(1)$&$1.4274863(4)$&$-0.0023078(4)$&$-0.0382833(1)$&$-0.0481719(1)$&$-0.04163115(7)$&$-0.02520187(3)$\\
  $\bar{E}_{\rm VMC}$ & $5.808195(6)$&$2.305629(3)$&$1.139746(1)$&$0.1238369(3)$&$-0.018672(1)$ &$-0.0528478(8)$&$-0.0613822(7)$&$-0.0509599(3)$ & $-0.0313879(1)$\\
  $\bar{E}_{\rm DMC}$ &$5.808184(7)$&$2.305612(4)$&$1.139729(2)$&$0.123820(2)$ &$-0.018689(1)$ &$-0.052860(1)$ &$-0.061395(1)$ &$-0.0509754(5)$ &$-0.0314033(3)$\\
  \hline
  Var. & $0.33(2)$&$0.102(6)$ & $0.060(3)$ & $0.0112(4)$& $0.0049(1)$& $0.00263(7)$& $0.00156(4)$& $0.000362(7)$& $0.000082(1)$\\
 \hline\hline \\[1em]
  \multicolumn{10}{c}{$N=259$} \\
 \hline \hline
 $r_\text{s}$ & 0.5 & 0.75 & 1.0 & 2.0 & 3.0 & 4.0 & 5.0 & 10.0 & 20.0 \\ 
 \hline
  $\bar{E}_{\rm HF}$  &$5.845614(2)$&$2.337769(1)$&$1.1685868(6)$&$1.4573846(1)$&$-0.0029783(1)$&$-0.03676961(8)$&$-0.04695788(6)$&$-0.04102113(2)$&$-0.02489611(1)$\\
  $\bar{E}_{\rm VMC}$ &$5.815060(3)$&$2.309840(2)$&$1.1426887(6)$&$0.1250068(6)$&$-0.0180072(5)$&$-0.0524023(3)$ &$-0.0610555(4)$ &$-0.0508438(2)$ &$-0.03134650(1)$\\
  $\bar{E}_{\rm DMC}$ &$5.815034(8)$&$2.309831(8)$&$1.142672(5)$ &$0.125000(1)$ &$-0.018016(1)$ &$-0.052418(1)$  &$-0.061065(1)$  &$-0.0508555(4)$ &$-0.0313595(5)$\\
  \hline
  Var. & $0.54(2)$ & $0.218(8)$ & $0.114(2)$ & $0.0233(6)$ & $0.0095(2)$ & $0.00526(7)$& $0.00308(4)$& $0.000682(8)$ & $0.000158(2)$ \\
 \hline\hline \\[1em]
 \multicolumn{10}{c}{$N=387$} \\
 \hline \hline
 $r_\text{s}$ & 0.5 & 0.75 & 1.0 & 2.0 & 3.0 & 4.0 & 5.0 & 10.0 & 20.0\\ 
 \hline
  $\bar{E}_{\rm HF}$  &$5.849115(1)$&$2.3401955(8)$&$1.1704411(5)$&$1.4669147(1)$&$0.0003433(1)$&$-0.03628659(6)$&$-0.04657043(4)$&$-0.04082637(1)$&$-0.024798478(7)$\\
  $\bar{E}_{\rm VMC}$ &$5.816893(3)$&$2.310961(2)$ &$1.143459(1)$ &$0.1253138(5)$&$-0.017831(1)$&$-0.0522918(6)$ &$-0.0609676(3)$ &$-0.0508112(2)$ & $-0.03133433(9)$ \\
  $\bar{E}_{\rm DMC}$ &$5.816883(8)$&$2.310942(5)$ &$1.143460(5)$ &$0.125301(2)$ &$-0.017848(2)$&$-0.052304(1)$  &$-0.060974(2)$  &$-0.0508228(2)$ & $-0.0313472(4)$\\
  \hline
  Var. & $0.77(4)$ & $0.32(1)$ & $ 0.177(6)$ & $0.0340(7)$ & $0.0136(2)$& $0.00730(9)$& $0.00460(5)$ & $0.00101(1)$ & $0.000228(2)$\\
 \hline\hline
 \end{tabular}
\end{table*}
\endgroup

Table \ref{table:EatInfi} shows the VMC and DMC energies at the infinite system size limit and the $\chi^2$ value for the fit of Eq.\ (\ref{eq:fs_extrap}). The increase in the $\chi^2$ value at high density is due to the crossover between HF-like behavior at small system sizes to interacting Fermi-liquid behavior at large system sizes. This behavior is much more pronounced in VMC than in DMC\@. 
\begin{table}[!htbp]
\centering
\caption{\label{table:EatInfi} VMC ($E_\text{VMC}$) and DMC ($E_\text{DMC}$) total energies of spin-polarized 3D-UELs at the infinite system size limit. The $\chi^2$ values for the fits of Eq.\ (\ref{eq:fs_extrap}) are reported. Also shown are the DMC energies of Spink \textit{et al.}\ \cite{Spink13}, which were obtained in TA calculations at a single finite cell size and include the analytical FS correction $\Delta E$.}
{\footnotesize
 \begin{tabular}{l c c c c c}
\hline\hline
 $r_\text{s}$ & $E_{\rm VMC}$ & $\chi^2_{\rm VMC}$ & $E_{\rm DMC}$ & $\chi^2_{\rm DMC}$ & Spink \textit{et al.}  \\ 
 \hline
 0.5  & $5.82045(7)$  & $235.22$ & $5.82043(6)$  & $34.46$ & $5.82498(2)$\\
 0.75 & $2.31313(4)$  & $186.85$ & $2.31314(1)$  & $3.12$  & \dots\\
 1.0  & $1.14497(1)$  & $72.07$  & $1.14498(2)$  & $14.22$ & $1.14634(2)$\\
 2.0  & $0.125920(5)$ & $66.03$  & $0.125912(1)$ & $0.14$  & $0.12629(3)$\\
 3.0  &$-0.017490(7)$ & $19.66$  &$-0.017497(3)$ & $2.38$  & $-0.017278(4)$\\
 4.0  &$-0.052059(4)$ & $27.48$  &$-0.052075(1)$ & $0.55$  &  \dots \\
 5.0  &$-0.060797(3)$ & $63.05$  &$-0.060806(4)$ & $3.79$  &  $-0.060717(5)$ \\
 10.0 &$-0.050753(2)$ & $67.92$  &$-0.050760(1)$ & $12.80$ & $-0.0507337(5)$ \\
 20.0 &$-0.031313(1)$ & $159.97$ &$-0.0313245(7)$& $1.94$  & $-0.0313160(4)$ \\ 
\hline\hline
 \end{tabular}
 }
\end{table}
Our DMC correlation energies are compared with previous results in Table~\ref{table:ECorr}. The results of Spink \textit{et al.}\ \cite{Spink13} were obtained using SJB-DMC without the ${\bm \pi}$ term in the BF\@. They corrected FS errors using canonical-ensemble twist averaging, and analytical corrections ($\Delta E$) \cite{Drummond08, Chiesa06}, but did not include the subsequently derived FS correction due to backflow \cite{Holzmann16}. To compare with the work of Spink \textit{et al.}\ \cite{Spink13}, we performed QMC calculations for $r_\text{s}=0.5$, 1, and 2 using a 118-electron fcc simulation cell, without including a ${\bm \pi}$ term in the BF\@. Our TA DMC energies for $r_\text{s}=0.5$, 1, and 2 are $5.81036(1)$, $1.14082(1)$, and $0.12436(2)$ Ha per electron, respectively, while the TA DMC energies reported by Spink \textit{et al.}\ \cite{Spink13} for the same system size and density without analytical FS correction are $5.80967(2)$, $1.14036(2)$, and $0.12403(3)$ Ha per electron, respectively. Our TA DMC energies agree with those of Spink \textit{et al.}\ \cite{Spink13} to within error bars. Hence the main source of difference between our DMC energies at the thermodynamic limit and the results of Spink \textit{et al.}\ energies is the difference between the two different approaches used for FS correction that are compared in Fig.\ \ref{fig:FS}. The density parameter interpolation (DPI) parameterization (Table~\ref{table:ECorr}) was obtained using four high-density and three low-density constraints on a seven-parameter correlation functional \cite{Sun10}. The DPI correlation energies are lower than the widely used correlation functionals in density functional calculations. 
\begin{table}[!htbp]
\centering
\caption{\label{table:ECorr} Correlation energy of the spin-polarized 3D-UEL at the infinite system-size limit. Energies are in mHa per electron. Our DMC correlation energies (pres.\ work) are compared with the DMC results of Spink \textit{et al.}\ \cite{Spink13}, the DMC results of Ceperley  \textit{et al.} \cite{Ceperley80}, and Ruggeri \textit{et al.}\ \cite{Ruggeri18}, which were calculated using a combination of FCIQMC and DMC. Density parameter interpolation (DPI) energies \cite{Sun10} are also listed\@. }
{\footnotesize
 \begin{tabular}{l c c c c c}
\hline\hline
 $r_\text{s}$ & Pres.\ work & Spink \textit{et al.} & Ruggeri \textit{et al.} & DPI & Ceperley \textit{et al.} \\ 
 \hline
 0.5  & $-41.06(7)$  & $-36.51(2)$ & $-40.44(5)$&$-40.91$&\dots   \\
 0.75 & $-35.41(1)$  & \dots       & \dots      &\dots   &\dots    \\
 1.0  & $-31.77(2)$  & $-30.41(2)$ & $-31.70(4)$&$-31.99$&\dots    \\
 2.0  & $-23.962(1)$ & $-23.58(3)$ & \dots      &$-23.6$ & $-24.0(3)$   \\
 3.0  & $-19.968(3)$ & $-19.749(4)$&  \dots     &\dots   &\dots    \\
 4.0  & $-17.387(1)$ & \dots       &  \dots     &\dots   &\dots     \\
 5.0  & $-15.515(4)$ & $-15.426(5)$& \dots      &$-15.1$ & $-15.4(1)$    \\
 10.0 & $-10.574(1)$ &$-10.5485(5)$& \dots      &$-10.2$ & $-10.5(1)$    \\
 20.0 & $-6.8469(7)$ & $-6.8384(4)$& \dots      &$-6.63$ & $-6.78(2)$    \\ 
\hline\hline
 \end{tabular}
    }
\end{table}

The correlation energy is defined as the difference between the Hartree-Fock energy per electron and the exact ground-state energy per electron, where the latter is approximated by our SJB-DMC results extrapolated to the limit of infinite system size. We fitted the DMC correlation energies to
\begin{equation}
    E_\text{c}(r_\text{s}) = \frac{A\ln(r_\text{s})+B+\gamma r_\text{s}}{1+\beta_1 r_\text{s}^{3/2}+\beta_2 r_\text{s}^2}, \label{eq:new_lda_fnal}
\end{equation}
which describes logarithmic behavior at very small $r_\text{s}$ \cite{Gell57}. This five-parameter fitting function gives a $\chi^2$ value of 5.8 per degree of freedom. The fitted values of $A$, $B$, and $\gamma$ are $0.0169245$, $-0.0250295$, and $-0.0174433$ Ha/electron with  asymptotic standard errors of $0.0003118$, $0.0004773$, and $0.001581$, respectively. The fitting parameters of $\beta_1$ and $\beta_2$ are $0.283806$ and $0.0520433$, respectively, with  asymptotic standard errors of $0.03059$ and $0.004198$, respectively. Equation (\ref{eq:new_lda_fnal}) has the form of the first two terms in the Gell-Mann-Bruckner expansion at small $r_\text{s}$ \cite{Gell57}, and it behaves like the Perdew-Zunger parameterization at large $r_\text{s}$ \cite{Perdew81}\@.

\textit{Conclusion.}  The ground state energy of the ferromagnetic 3D-UEL within the density range $0.5 \leq r_\text{s} \leq 20 $ was calculated using the VMC and DMC methods. The single-particle and many-body FS errors were corrected by canonical-ensemble twist averaging and extrapolation to the thermodynamic limit. Our correlation energies are more negative than previous works so the variational principle suggests that they are more accurate, and we have used them to parameterize a correlation functional.

\textit{Acknowledgments.} S.A.\ and S.M.V.\ acknowledge support from the UK EPSRC grants EP/P015794/1 and EP/W010097/1, and the Royal Society. S.A. gratefully acknowledges the generous allocation of computing time by the Paderborn Center for Parallel Computing (PC2) on the FPGA-based supercomputer NOCTUA2 through project TMOMC. 

\bibliography{FerroHEG}

\end{document}